\documentstyle[twoside,fleqn,espcrc2,epsf]{article}

\title{The Low-lying Dirac Eigenmodes from Domain Wall Fermions}

\author{
 Guofeng Liu\address{Department of Physics, Columbia University,
 New York, NY, 10027, USA }\thanks{ 
 This work was supported in part by the Department of Energy and the
 RIKEN BNL Research Center.
 } 
 [RBC colaboration]
}

\newcommand{\be}{\begin{equation}}
\newcommand{\ee}{\end{equation}}

\newcommand{\bd}{\begin{displaymath}}
\newcommand{\ed}{\end{displaymath}}

\newcommand{\bea}{\begin{eqnarray}}
\newcommand{\eea}{\end{eqnarray}}

\newcommand{\ba}[1]{\begin{array}{#1}}
\newcommand{\ea}{\end{array}}

\def\lvec#1{\setbox0=\hbox{$#1$}
    \setbox1=\hbox{$\scriptstyle\leftarrow$}
    #1\kern-\wd0\smash{
    \raise\ht0\hbox{$\raise1pt\hbox{$\scriptstyle\leftarrow$}$}}
    \kern-\wd1\kern\wd0}
\def\rvec#1{\setbox0=\hbox{$#1$}
    \setbox1=\hbox{$\scriptstyle\rightarrow$}
    #1\kern-\wd0\smash{
    \raise\ht0\hbox{$\raise1pt\hbox{$\scriptstyle\rightarrow$}$}}
    \kern-\wd1\kern\wd0}

\newcommand{\beq}{\begin{equation}}
\newcommand{\eeq}{\end{equation}}

\newcommand{\pslash}[1]{\rlap{/}\kern-0.8pt #1}
\newcommand{\lslash}{\rlap{/}\kern-0.0pt l}
\newcommand{\Dslash}{\rlap{/}\kern-2.0pt D}

\def\qbq{\langle \overline{q} q \rangle}

\def\dmi{\delta m_i}

\def\mpi2{m_\pi^2}

\def\qw{Q^{\rm (w)}}

\begin{document}

\begin{abstract}
We calculate the low-lying eigenvalues and eigenvectors of the hermitian domain 
wall Dirac operator on various gauge backgrounds by Ritz minimization. The mass 
dependence of these eigenvalues is studied to extract the physical 4 
dimensional $\lambda$, whose spectral density is related to $\langle \bar{\psi} \psi \rangle$ 
through the Banks-Casher relation, and $\delta m$, which represents the effects 
of the residual chiral symmetry breaking in domain wall formalism on a per
eigenmode basis. The topological structure of the underlying gauge field is 
examined by measuring the $\Gamma_5$ matrix elements between the low-lying 
eigenmodes. 
\end{abstract}
\maketitle
\section{DIAGONALIZATION METHOD }
We use the conjugate-gradient method proposed by Kaukreuter and Simma
~\cite{Kalkreuter}\footnote{QCDSP implementation by Robert Edwards.} to 
calculate the lowest 19 eigenvalues and eigenvectors of the hermitian domain 
wall Dirac operator $D_H=\gamma_5 R D$, where $D$ is the domain wall Dirac
operator and $R$ is the reflection operator in $s$ direction (look at
\cite{blum2000} for a detailed description of the conventions). 
\begin{table}[htb] 
\vskip -0.3in
\caption{
  The bare quark masses used for the Wilson ensembles are 
  0.0, 0.0025, 0.005, 0.0075, 0.01. 
  For the Iwasaki ensemble, only the lowest mass, 0.0005, is different.}
\label{config_list}
\begin{tabular} {cccccc}
\hline
\hline
V & Action & $\beta$ & $a^{-1}/$Gev & $L_s$ & \#conf.   \\
\hline
$16^4$         & Wilson  & 6.0  & 2.0 & 16 & 32 \\
$16^4$         & Wilson  & 6.0  & 2.0 & 8  & 10 \\
$16^4$         & Iwasaki & 2.6  & 2.0 & 16 & 55 \\
\hline
\end{tabular}
\vskip -0.3in
\end{table}
\par
For each of the configurations listed in Table~\ref{config_list}., we calculate 
the eigenvalue spectrum at five different masses. There are two main reasons 
for this practice. First, by studying the mass dependence as described in 
section 2, we can remove from the eigenvalue spectrum a portion of the residual 
chiral symmetry breaking effects. For the Iwasaki action at 
$\beta=2.6$, the residual mass is very small, which makes the convergence 
of the Ritz minimization prohibitively slow at zero bare quark mass. 
Thus, we had to use nonzero bare quark masses ($5\times 10^{-4}$ and up) 
to speed up the convergence, and then extrapolate to the region of interest. 
\section{SPECTRAL DENSITY AND CHIRAL SYMMETRY BREAKING} 
In the continuum limit, the spectral density of Dirac operator is related to 
the chiral condensate $\langle \bar{\psi} \psi \rangle$ by the Banks-Casher 
relation 
\begin{eqnarray}
  -\langle \bar{\psi} \psi \rangle
    =  \frac{1}{12V} \frac { \langle | \nu | \rangle }{m} + \frac{1}{12V} \left \langle \sum_{\lambda \ne 0} \frac{m}{\lambda^2 + m^2} \right \rangle
   \label{eq:banks_casher_lat4} 
\end{eqnarray}
and 
\begin{eqnarray}
 \lim_{m\rightarrow 0}\lim_{V\rightarrow \infty} -\langle \bar{\psi} \psi \rangle  &=& {\pi \over 12} \rho(0). \label{eq:banks_rho0}
\end{eqnarray}
Since $D_H$ is a continuous function of $m_f$, we can expand and
parameterize its eigenvalue $\Lambda_{H,i}$ as  
\begin{equation}
   \Lambda_{H,i}^2 = n_{5,i}^2( \lambda^2_i + ( m_f + \dmi)^2) + O(m_f^3).
   \label{eq:dherm_lambda_vs_mf}
\end{equation}
From 
\begin{eqnarray}
  {dD_H\over dm_f}= -\gamma_5 \qw, 
\end{eqnarray}
where $-m_fR\qw$ is the mass term in the fermion matrix, we can derive
\begin{eqnarray}
     - \langle \Lambda_{H,i}|\gamma_5\qw|\Lambda_{H,i} \rangle 
     &=&{d\Lambda_{H,i}\over dm_f} \\
     &=& { n^2_{5,i}(m_f+\delta m_i)\over \Lambda_{H,i} } 
\end{eqnarray}
The quantity $\langle \bar{\psi} \psi \rangle$ is defined on the boundary for 
domain wall fermions, and given by 
\begin{eqnarray}
  -\qbq 
    & = & -\frac{1}{12 V}  \left\langle \sum_{\Lambda_H} 
        \frac{ \langle \Lambda_H | \gamma_5 \qw | \Lambda_H
        \rangle } { \Lambda_H} \right\rangle \label{eq:dh_spectral_qbq} \\
    & = & {1\over 12 V}\left \langle \sum_i \frac{m_f + \dmi}
    {\lambda_i^2 + ( m_f + \dmi)^2} \right \rangle.
  \label{eq:dwf_banks_casher}
\end{eqnarray}
By comparing Eq.~\ref{eq:dwf_banks_casher} and Eq.~\ref{eq:banks_casher_lat4},
we can recogonize the parameter $\lambda_i$ as an eigenvalue of the four
dimensional Dirac operator. Here, $\delta m_i$ represents the contribution to 
the eigenvalue from the chiral symmetry breaking effects of coupling of the
domain walls. 
\begin{figure}[htb]
\epsfxsize=\hsize
\vskip -0.1in
\epsfbox{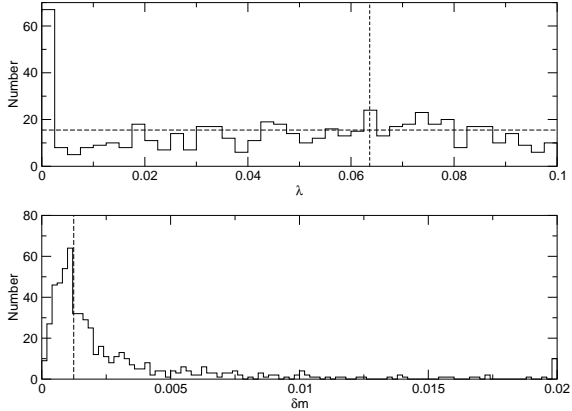}
\vskip -0.4in
\caption{$\lambda$ and $\delta m$ distribution for the Wilson ensemble at 
$\beta=6.0$, $L_s=16$. The horizontal line in the upper panel is $\rho(0)$
calculated from $\langle \bar{\psi} \psi \rangle$ using 
Eq.~\ref{eq:banks_rho0}. The vertical
line in the upper panel is the minimum of the largest $\lambda$ in each 
configuration, which means there is no error introduced by dropping higher
modes to the left of this line. The vertial line in the lower panel is the 
residual mass determined independently from the midpoint Ward/Takahashi 
identity term.} 
\label{fig:hist_16nt16_ls16_b6.0_wilson_m1.8}
\vskip -0.2in
\end{figure}
\begin{figure}[htb]
\epsfxsize=\hsize
\epsfbox{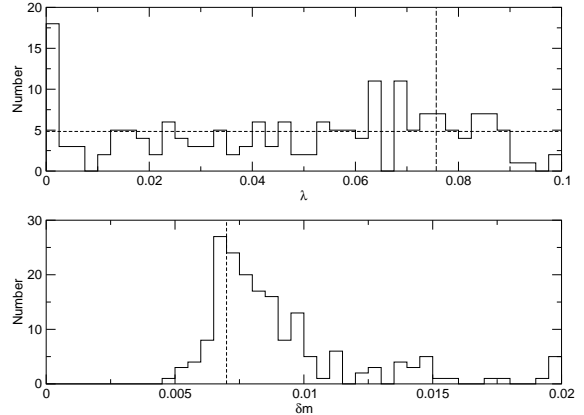}
\vskip -0.4in
\caption{$\lambda$ and $\delta m$ distribution for Wilson ensemble
  at $\beta=6.0$, $L_s=8$.}
\label{fig:hist_16nt16_ls8_b6.0_wilson_m1.8}
\end{figure}
\begin{figure}[htb]
\epsfxsize=\hsize
\vskip -0.2in
\epsfbox{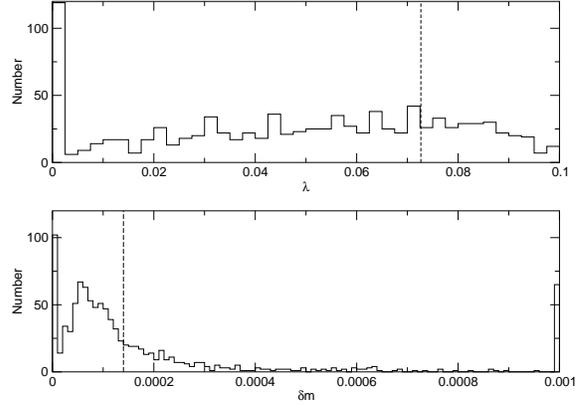}
\vskip -0.4in
\caption{$\lambda$ and $\delta m$ distribution for Iwasaki ensemble 
  at $\beta=2.6$, $L_s=16$.} 
\label{fig:hist_16nt16_ls16_b2.6_Iwasaki_m1.8}
\vskip -0.3in
\end{figure}
\par
Fig.~\ref{fig:hist_16nt16_ls16_b6.0_wilson_m1.8},
\ref{fig:hist_16nt16_ls8_b6.0_wilson_m1.8} and 
\ref{fig:hist_16nt16_ls16_b2.6_Iwasaki_m1.8} are the distribution of $\lambda$
and $\delta m$ for the ensembles listed in Table~\ref{config_list}. A rough
agreement between the peak of the $\delta m$ distribution and $m_{res}$ from
the mid-point term is observed. A nice agreement between the spectral
density and $\langle \bar{\psi} \psi \rangle$ is confirmed except for 
the peak at $\lambda = 0$, which is explained by the abundance of
zero modes at finite volume, and is expected to disappear in the infinite
volume limit. 
The comparison between Fig~\ref{fig:hist_16nt16_ls16_b6.0_wilson_m1.8} and 
\ref{fig:hist_16nt16_ls8_b6.0_wilson_m1.8} shows the expected $L_s$
dependence of $\delta m$. 
Fig~\ref{fig:hist_16nt16_ls16_b6.0_wilson_m1.8}
and \ref{fig:hist_16nt16_ls16_b2.6_Iwasaki_m1.8} 
have the same $L_s$, volume and strength of coupling but different types of 
action, which makes the $\delta_m$ an order of magnitude smaller for the
latter.
%
\section{GAUGE FIELD TOPOLOGICAL STRUCTURE}
In the continuum limit, the Dirac operator has $\gamma_5$ symmetry, which
ensures that the zero modes have chirality of $\pm 1$, and the non-zero modes, 
with chirality $0$, are paired by the application of $\gamma_5$. The number 
of zero modes should correspond to the winding number of the gauge field. 
\begin{figure}[htb]
\vskip -0.1in
\epsfxsize=\hsize
\epsfbox{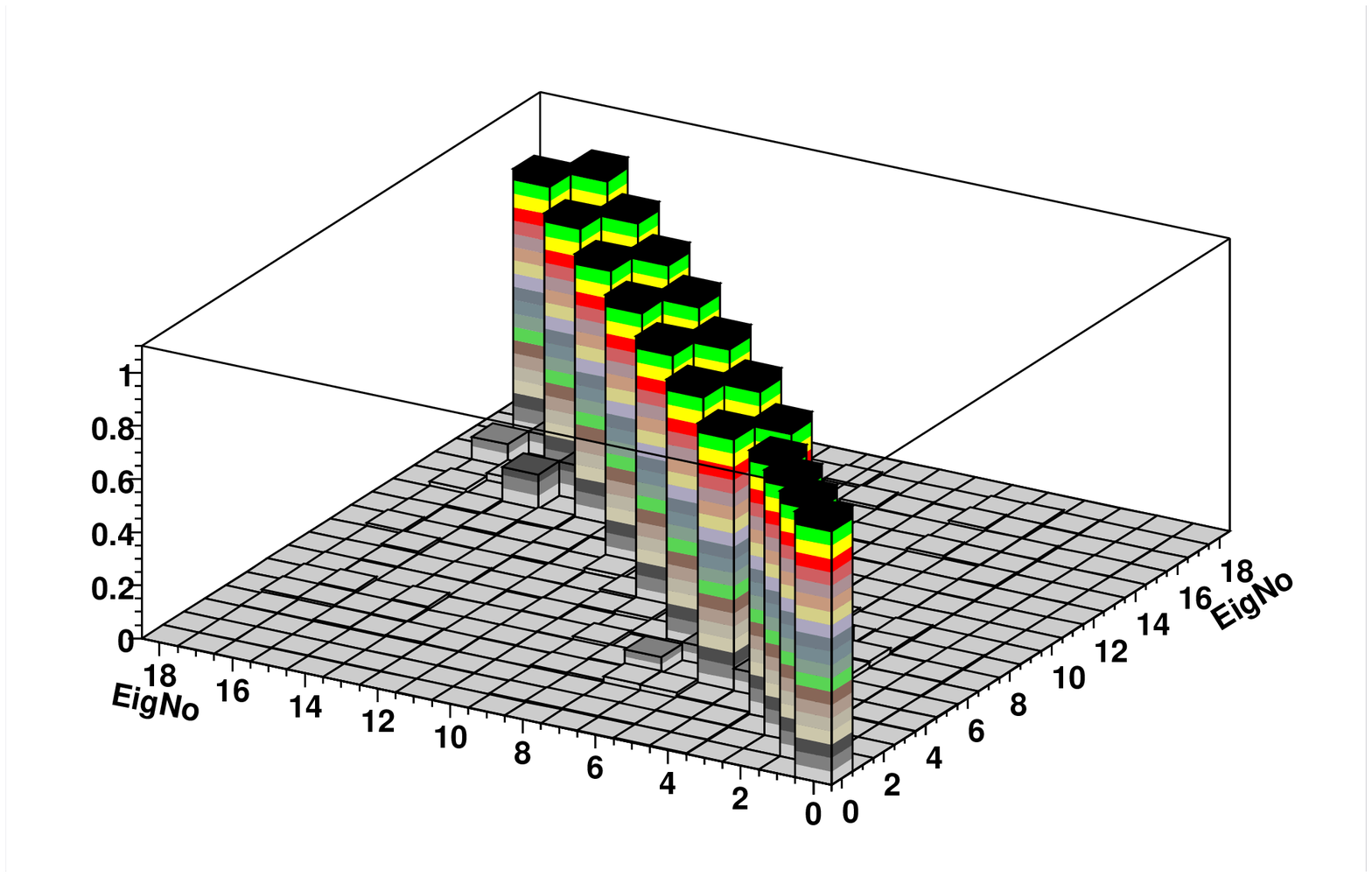}
\nonumber
\end{figure}
\begin{figure}[htb]
\vskip -0.53in
\epsfxsize=\hsize
\epsfbox{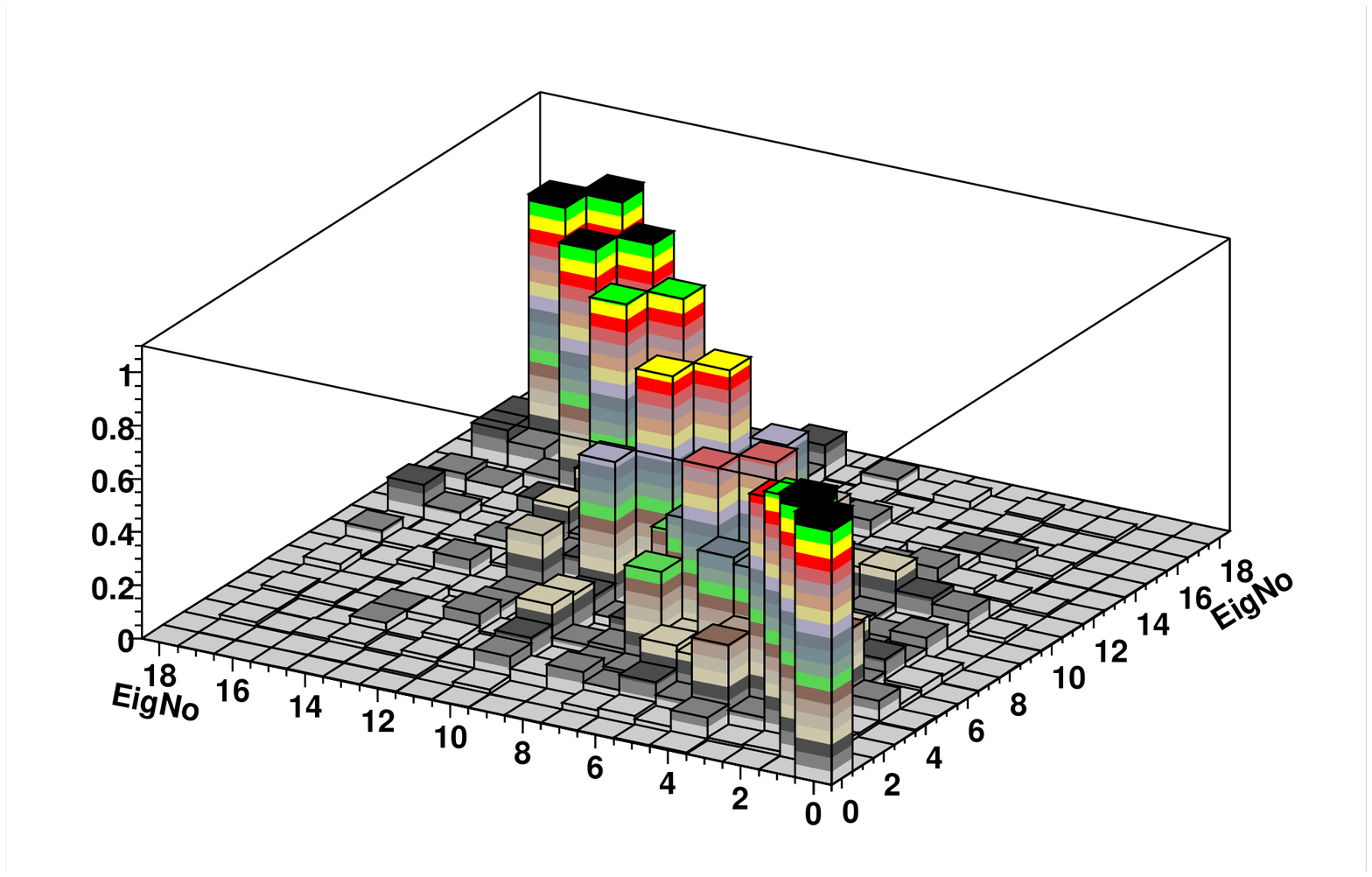}
\nonumber
\end{figure}
\begin{figure}[htb]
\vskip -0.55in
\epsfxsize=\hsize
\epsfbox{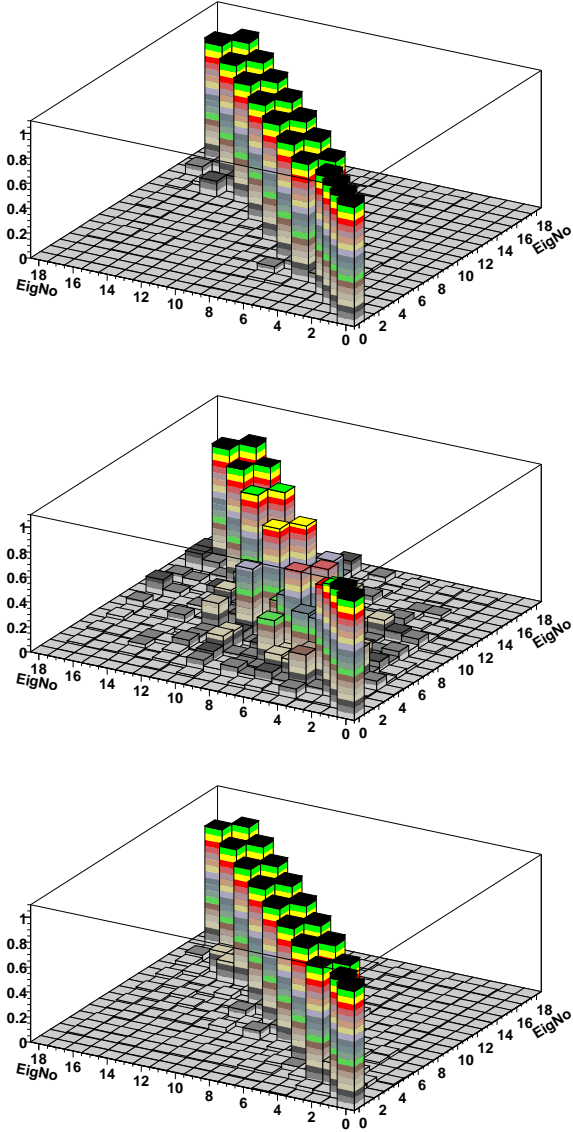}
\vskip -0.4in
\caption{Lego plots of three sequential configurations separated by 3 and 4 
  heat bath sweeps. } 
\vskip -0.3in
\label{fig:lego}
\end{figure}
\par
In the domain wall formalism, we display the matrix elements 
  $|\langle \Lambda_{H,i} | \Gamma_5 | \Lambda_{H,j} \rangle |$ as a 
3 dimensional lego plot. 
This provides a way of visualizing the chiral properties of the domain wall
Dirac operator and the topological structure of the underlying gauge field. 
Fig~\ref{fig:lego} shows a set of such lego plots for the Wilson ensemble at
$\beta=6.0$, $L_s=16$. In the first lego plot, the lowest 4 of the  
eigenvectors are also $\Gamma_5$ eigenvectors, while the rest of them are 
paired. Fig~\ref{fig:lego} also shows that during the heat bath evolution, 
when a configuration transforms from the topological sector with topology 4 
to that with topology 2, some complex configurations result in between 
which can not be unambiguously categorized as belonging to any regular 
topological sector. Table~\ref{complex_list} lists the percentage of this kind 
of complex configuration for each ensemble in Table~\ref{config_list}. 
The readers are refered to C. Dawson's poster for a more quantative study
of the complex configurations.
\begin{table}[htb]
\vskip -0.3in
\caption{List of the values of $m_{res}$ and percentage of complex
configurations.}
\label{complex_list}
\begin{tabular} {ccccc}
\hline
\hline
Action & $\beta$ & $L_s$ & $m_{res}$ & complex\%\\
\hline
Wilson  & 6.0  & 16 & 0.00124(5) & 50\% \\
Wilson  & 6.0  & 8  & $\sim0.007$& 50\%\\
Iwasaki & 2.6  & 16 & 0.00014(3) & 10\%\\
\hline
\end{tabular}
\vskip -0.3in
\end{table}
%
\section{CONCLUSIONS}
Our prescription of extracting $\lambda$ and $\delta m$ distribution 
from mass dependence of the eigenvalues of $D_H$ is consistent with the 
Banks-Casher relation and the usual ways of determing $\langle\bar{\psi}\psi
\rangle$ and $m_{res}$. The good chiral properties of domain wall fermions 
makes them a very good tool for studying the topological structure of the 
background gauge field. We find that this structure is closely related to the
type of action used in generating these configurations. 
\section{ACKNOWLEDGMENT}
These calculations were performed on the QCDSP machines at Columbia
and the RIKEN BNL Research Center. 

\end{document}